\documentclass[conference, 10pt]{IEEEtran}

\usepackage{cite}
\usepackage{url}
\usepackage{graphicx}
\usepackage{color}
\usepackage{placeins}
\usepackage{float}
\usepackage{tabularx,colortbl}
\usepackage{ifthen}

\usepackage{balance}
\usepackage{hft-thesis}

\RequirePackage[caption=false,font=footnotesize]{subfig}

\makeatletter

\newcounter{author}
\renewcommand{\author}[2][]{
   \stepcounter{author}
   \@namedef{author@\theauthor}{#2}
   \@namedef{authorlabel@\theauthor}{#1}
}

\newcounter{address}
\newcommand{\address}[2][]{
   \stepcounter{address}
   \@namedef{address@\theaddress}{#2}
   \@namedef{addresslabel@\theaddress}{#1}
}

\newcommand{\alsep}{and}

\def\newmaketitle{\par%
  \begingroup%
  \normalfont%
  \def\thefootnote{}%  the \thanks{} mark type is empty
  \def\footnotemark{}% and kill space from \thanks within author
  \let\@makefnmark\relax% V1.7, must *really* kill footnotemark to remove all \textsuperscript spacing as well.
  \footnotesize%       equal spacing between thanks lines
  \footnotesep 0.7\baselineskip%see global setting of \footnotesep for more info
  \normalsize%
  \twocolumn[\thenewmaketitle\@IEEEaftertitletext]%
  \if@IEEEusingpubid
     \enlargethispage{-\@IEEEpubidpullup}%
  \fi
  \endgroup
  \setcounter{footnote}{0}\let\maketitle\relax\let\@maketitle\relax
  \gdef\@thanks{}%
  \let\thanks\relax}

\def\thenewmaketitle{
  \newpage
  \begin{center}%
    \vskip0.2em{\Huge\@IEEEcompsoconly{\sffamily}\@IEEEcompsocconfonly{\normalfont\normalsize\vskip 2\@IEEEnormalsizeunitybaselineskip
   \bfseries\large}\@title\par}\vskip1.0em\par%
    \vspace{1ex}
    \newcounter{c@author}
    \newcounter{c@tmp}
    \ifthenelse{\value{author}=2}{%
      \newcommand{\liand}{ and }}{%
      \newcommand{\liand}{, and }}
    \ifthenelse{\value{address}<2}{%
      \@nameuse{author@1}%
      \stepcounter{c@author}%
      \whiledo{\value{c@author}<\value{author}}{%
        \setcounter{c@tmp}{\value{author}}%
        \addtocounter{c@tmp}{-\value{c@author}}%
        \ifthenelse{\value{c@tmp}=1}{%
          \renewcommand{\alsep}{\liand}}{\renewcommand{\alsep}{, }}%
        \stepcounter{c@author}\alsep \@nameuse{author@\thec@author}}\\%
    }
    {%Add address references after the author's name
      \@nameuse{author@1}${}^{(\ref{\@nameuse{authorlabel@1}})}$%
      \stepcounter{c@author}%
      \whiledo{\value{c@author}<\value{author}}{%
      \setcounter{c@tmp}{\value{author}}%
      \addtocounter{c@tmp}{-\value{c@author}}%
      \ifthenelse{\value{c@tmp}=1}{%
        \renewcommand{\alsep}{\liand}}{\renewcommand{\alsep}{, }}%
      \stepcounter{c@author}\alsep \@nameuse{author@\thec@author}%
        ${}^{(\ref{\@nameuse{authorlabel@\thec@author}})}$%
      }
    }
    \vspace{0.2ex}

    \ifthenelse{\value{address}>0}{%
      \ifthenelse{\value{address}=1}{
        {\@nameuse{address@1}}
      }
      {%Output the addresses as an enumerated list
        \newcounter{c@address}

        \begin{center}
        \whiledo{\value{c@address}<\value{address}}
        {
          \refstepcounter{c@address}
            ${}^{(\thec@address)}$\,%
              \label{\@nameuse{addresslabel@\thec@address}}%
              \@nameuse{address@\thec@address}\\ %
        }
        \end{center}
      } % end of the address creation ifthenelse block
    }
    {
      \relax
    }
  \end{center}
}

\makeatother

\title{3-D Near-Field Passive Radar Imaging Using Multiple Illumination Sources}

\author[org1]{Quanfeng Wang}
\author[org2]{Mei Song Tong}
\author[org1]{Thomas F. Eibert}

\address[org1]{Department of Electrical Engineering, School of Computation, Information and Technology,\\
Technical University of Munich, Munich, Germany, quanfeng.wang@tum.de}
\address[org2]{{Department of Electronic Science and Technology, Tongji University, Shanghai, China}}

\AddToHook{shipout/foreground}{
  \put(0.5\paperwidth, -0.8cm){ 
    \makebox[0pt][c]{
      \begin{minipage}{\textwidth}
        \centering \scriptsize 
          This article has been accepted for publication in 2025 IEEE International Symposium on Antennas and Propagation and North American Radio Science Meeting (AP-S/CNC-USNC-URSI). This is the author's version which has not been fully edited and content may change prior to final publication.
          Citation information: 10.1109/AP-S/CNC-USNC-URSI55537.2025.11266162
      \end{minipage}
    }
  }
}

\AddToHookNext{shipout/foreground}{
  \put(0.5\paperwidth, -\paperheight + 0.7cm){ 
    \makebox[0pt][c]{
      \begin{minipage}{\textwidth}
        \centering \scriptsize
          Copyright~\copyright~2025 IEEE. Personal use of this material is permitted. Permission from IEEE must be obtained for all other uses, in any current or future media, including\\reprinting/republishing this material for advertising or promotional purposes, creating new collective works, for resale or redistribution to servers or lists, or reuse of any copyrighted component of this work in other works by sending a request to pubs-permissions@ieee.org.
      \end{minipage}
    }
  }
}

\begin{document}

\newmaketitle

\begin{abstract}
Near-field (NF) passive radar imaging depends on the illumination of the imaging scene by a non-cooperative transmitter (Tx). It is demonstrated that combining imaging results obtained with Tx antennas at different positions can enhance the performance of passive radar imaging. On the one hand, multiple Tx antennas provide diverse illumination perspectives, reducing the likelihood of unilluminated regions on the targets of interest (TOIs). On the other hand, the coherent summation of imaging results obtained for different illuminations helps to suppress potential artifacts. This approach is in particular advantageous for imaging complex objects with concave structures such as dihedral arrangements, where the ghosts due to multiple reflections are highly configuration-dependent. For each illuminating configuration, a single-frequency inverse source solver is utilized to reconstruct the equivalent sources of the TOIs and the resulting single-frequency images are then superimposed coherently with corresponding phase and magnitude correction methods. The obtained multi-frequency images are finally coherently combined to enhance the imaging quality. Both simulation and measurement results are presented to validate the effectiveness of the approach.

\end{abstract}

\section{Introduction}
Microwave imaging has garnered considerable interest over the past decades in fields such as security inspections~\cite{sheen2001threedimensionala} and medical diagnostics~\cite{ahmed2012advanced}. Active imaging systems such as monostatic synthetic aperture radar (SAR) require repeated measurements for transmitter (Tx) and receiver (Rx) pairs at different positions~\cite{lopez-sanchez20003d}. Similarly, multiple-input-multiple-output SAR imaging setups rely on cooperative scanning systems consisting of several Tx and Rx antennas~\cite{zhuge2012threedimensional}. In contrast to active microwave imaging, the concept of passive radar imaging, which utilizes ubiquitous radiation and does not rely on any active scanning system, has been applied to various infrastructures, such as global navigation satellite systems (GNSSs) in the far-field (FF) region~\cite{pastina2021passive} and WiFi in the near-field (NF) region~\cite{holl2017holography}. Although radio tomographic imaging utilizing  multiple WiFi transceiver nodes surrounding the domain of interest has been demonstrated~\cite{dubey2024reconciling}, most passive radar imaging approaches still focus on a single illuminating Tx source. With the widespread use of wireless communication technologies, multiple Tx antennas in NF environments, e.g., various WiFi access points, become more and more commmon and leveraging these multiple Tx sources to enhance imaging performance is, thus, an attractive option.

In terms of accurate image reconstruction and object recognition, the multiple reflections that exist in highly reflective environments or concave structures often pose a significant challenge. Efforts have been undertaken to address this issue. For instance, an array rotation technique has been applied to suppress multipath ghosts, based on the observation that the ghosts depend on the scanning configuration, whereas actual targets are configuration-independent~\cite{guo2018multipath}. Circular polarization measurements have been utilized to separate and compensate for multipath reflections~\cite{si2021accurate}. In addition, a ray-tracing based phase correction method has been proposed for dihedral-angle objects~\cite{liang2024investigations}. However, these methods are primarily designed for active imaging systems and may be constrained by the complex adaptation of the measurement array~\cite{guo2018multipath,si2021accurate}, or limited to specific configurations of monostatic antenna arrays and objects~\cite{liang2024investigations}.

An inverse source reconstruction based 3-D NF passive radar imaging technique is presented in~\cite{wang2024TAP}, where the equivalent sources in the form of plane-wave spectra (PWS) are reconstructed using a single-frequency inverse source solver. The single-frequency images are generated from the reconstructed PWS and coherently superimposed to create the final multi-frequency image with corresponding phase and magnitude corrections. However, the presented imaging mechanism considers only a single Tx antenna. Moreover, the proposed phase correction method accounts only for the direct propagation path from the Tx antenna to the TOIs and to the measurement probe, without considering the contributions of multiple interactions. Consequently, it may also suffer from multipath ghosts when imaging TOIs with strong multiple interactions, such as dihedral structures. In this work, the combination of multiple illumination sources is considered, and it is demonstrated that this approach can enhance imaging performance and suppress multipath ghosts in such scenarios.

In Section~\ref{sec:img_gen}, the imaging algorithm for combining contributions from multiple Tx sources is introduced. Both simulation and measurement results are presented in Section~\ref{sec:numerical}. Conclusions are drawn in Section~\ref{sec:conclusion}.

\section{Imaging Algorithm}
\label{sec:img_gen}
\begin{figure}[t]
	\centering
	\includegraphics[scale=0.65]{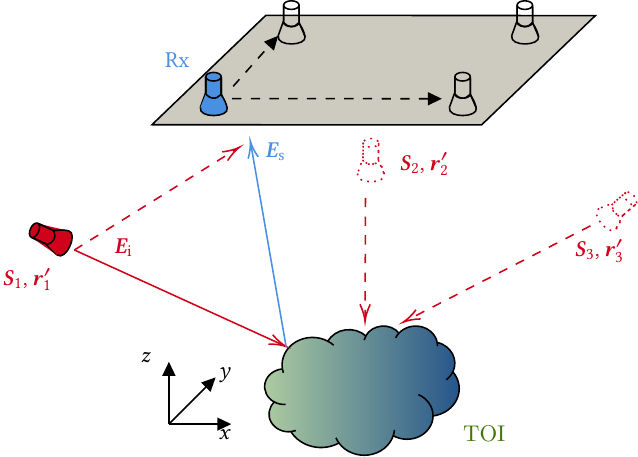}
	\caption{Imaging configuration of the TOI with a single Tx antenna, which moves and is excited sequentially from various locations.}
	\label{fig:configuration}
\end{figure}%
The considered imaging configuration is illustrated in Fig.~\ref{fig:configuration}. Illumination can be achieved using multiple consecutively excited Tx antennas, or, as assumed in this work, a single Tx antenna that moves and illuminates from different locations. Planar measurements comprising a sufficient number of sampled scattered fields are carried out using a single scanning probe for each illumination. The measured signal of the probe at observation position $\veg{r}_{m}$, under the illumination of the Tx source $\veg{S}_{n}$ located at $\veg{r}'_{n}$, is expressed as (a time dependency of $\e^{\jm \omega t}$ is suppressed)
\begin{align}
  U\Big(\veg{S}_{n},\veg{r}_{m}\Big)=&\iiint_{V_{\rm w}}\veg{w}(\veg{r}-\veg{r}_{m})\cdot\bigg[\veg{E}_{\rm i}\Big({\veg r},\veg{J}_{\rm i}\left(\veg{S}_{n}\right)\Big)  \notag\\
&+\veg{E}_{\rm s}\Big( {\veg r},\veg{J}_{\rm s}\left(\veg{S}_{n}\right)\Big) + \veg{E}_{\rm sp}\Big({\veg r},\veg{J}_{\rm sp}\left(\veg{S}_{n}\right)\Big)\bigg] \dd v\,,\label{eq:Urm}
\end{align}
where $\veg{w}(\veg{r}-\veg{r}_{m})$ describes the receiving behavior of the probe with volume $V_{\rm w}$ at the observation location $\veg{r}_{m}$. $\veg{J}_{\rm i}\left(\veg{S}_{n}\right)$, $\veg{J}_{\rm s}\left(\veg{S}_{n}\right)$, and $\veg{J}_{\rm sp}\left(\veg{S}_{n}\right)$ represent the equivalent sources of the Tx antenna, the TOIs, and of potential parasitic echoes, respectively. The equivalent sources are reconstructed in the form of PWS using the inverse source solver and are subsequently utilized to generate images for each discrete frequency \mbox{$f=1,\dots,F$}, and for \mbox{$n=1,\dots,N$} different illumination positions $\veg{r}'_{n}$ of the Tx antenna. Once the single-frequency images $\mathring{J}_{\mathrm{s},\,p}\left(k_{f},\veg{S}_{n}, \veg{r}'\right)$ have been obtained, they are superimposed coherently according to
\begin{align}
  J_{\mathrm{s},\,p}(\veg{r}')=\sum_{n=1}^{N}\sum_{f=1}^{F}&\psi_{\mathrm{s},\,p}\left(k_f,\veg{S}_{n},\veg{r}'\right) \notag \\
  &{\mathcal{M}_{\mathrm{ s},\,p}\left(k_f,\veg{S}_{n},\veg{r}'\right)}\,k_f\,\Delta k_f \mathring{J}_{\mathrm{s},\,p}\left(k_{f},\veg{S}_{n}, \veg{r}'\right) \,,\label{eq:sumF}
\end{align}
where $p\in\left\{x,y,z\right\}$ indicates three vector components in Cartesian coordinates. The phase and magnitude correction terms $\psi_{\mathrm{s},\,p}\left(k_f,\veg{S}_{n},\veg{r}'\right)$ and $\mathcal{M}_{\mathrm{ s},\,p}\left(k_f,\veg{S}_{n},\veg{r}'\right)$ are found in~\cite{wang2024TAP}.

\section{Results}
\label{sec:numerical}
% \begin{figure}[t]
% 	\centering
% 	\includegraphics[scale=0.45]{Figures/Pyramid.pdf}
% 	\caption{Illustration of the simulation setup with a pyramid-shaped convex structure located at the origin. Simulations were performed separately for two different positions $\veg{r}'_{1}$ and $\veg{r}'_{2}$ of the Hertzian dipole serving as the Tx source.}
% 	\label{fig:Pyramid}
% \end{figure}%
\begin{figure}[t]
	\centering
	\subfloat[]{\includegraphics[scale=0.27]{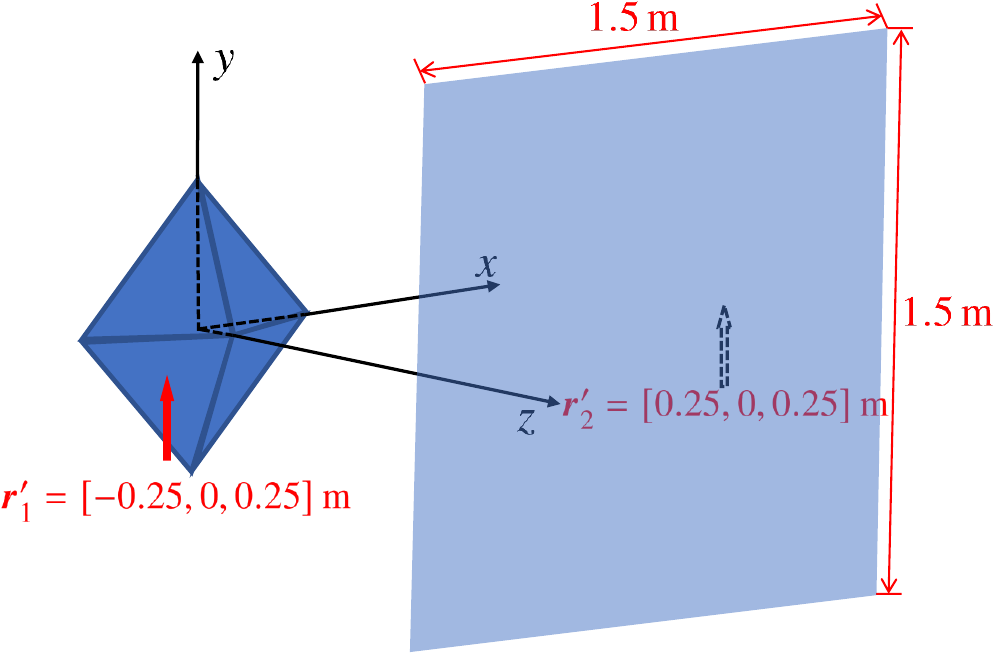}}%
	\hfill
	\subfloat[]{\includegraphics[scale=0.27]{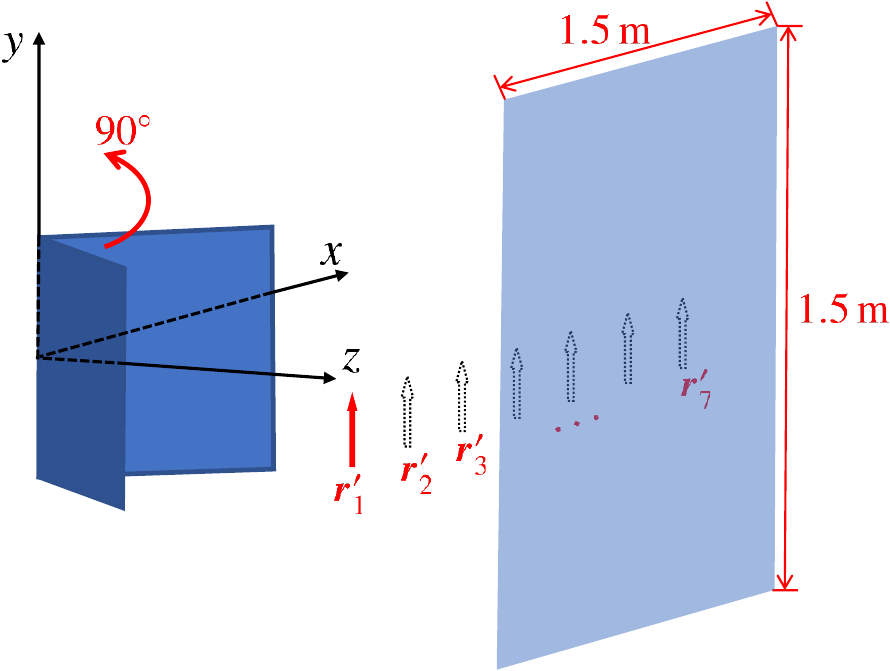}}%
	\caption{Illustration of the simulation setup with the TOIs located around the origin. Simulations were performed separately for different positions $\veg{r}'_{n}$ of the Hertzian dipole serving as the Tx source. The TOIs are 
  a pyramid-shaped convex structure~(a), and 
  a $90^{\circ}$ dihedral corner reflector~(b). }
	\label{fig:Simu}
\end{figure}

Simulations were conducted by the commercial software FEKO~\cite{EMSS2024}. The first simulation involved a perfectly electrically conducting (PEC) pyramid-shaped convex structure placed at the origin. A $y$-polarized Hertzian dipole serving as the illumination source was first placed at $\veg{r}'_{1}=[-0.25, 0, 0.25]\,$m and then at $\veg{r}'_{2}=[0.25, 0, 0.25]\,$m, as shown in Fig.~\ref{fig:Simu}(a). In both cases, the scattered fields consisting of the $x$- and $y$-components were collected at $10\,201$ uniformly distributed positions over a rectangular measurement plane at $z=\SI{1}{\meter}$, extending from $[x,y]=[-0.75,-0.75]\, $m to $[x,y]=[0.75,0.75]\, $m.  A total of 21 frequencies spanning from $\SI{6}{\giga\hertz}$ to $\SI{10}{\giga\hertz}$ were chosen with an equal step size of $\SI{200}{\mega\hertz}$. All of the given imaging results show the normalized vector magnitude intensity according to (\ref{eq:sumF}).
\begin{figure}[t]
	\centering
	\subfloat[]{\includegraphics[scale=0.62]{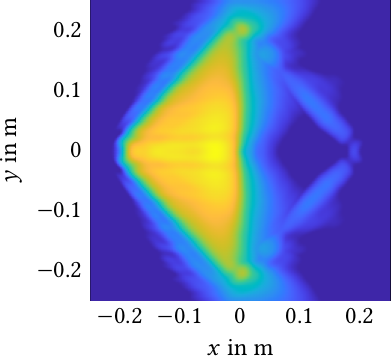}}%
	\hfill
	\subfloat[]{\includegraphics[scale=0.62]{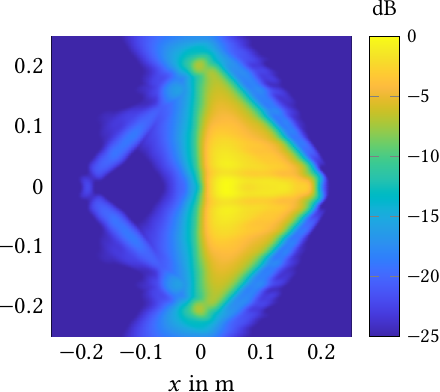}}%
	\caption{Imaging results of the pyramid-shaped object with the single illumination source located at $\veg{r}'_{1}=[-0.25, 0, 0.25]\,$m~(a), and $\veg{r}'_{2}=[0.25, 0, 0.25]\,$m~(b). }
	\label{fig:pyramid_sep}
\end{figure}

\begin{figure}[t]
	\centering
	\includegraphics[scale=0.8]{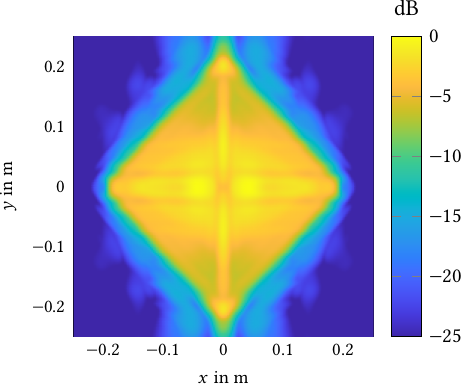}
	\caption{Imaging result of the pyramid-shaped object combining the illuminating contributions from the Tx antenna at $\veg{r}'_{1}$ and $\veg{r}'_{2}$.}
	\label{fig:pyramid_comb}
\end{figure}%

When only a single illumination source is involved, the corresponding imaging results displayed as 2-D maximum intensity projections (MIPs) are shown in Figs.~\ref{fig:pyramid_sep}(a) and (b), respectively. It is seen that neither the Tx source at $\veg{r}'_{1}$ nor at $\veg{r}'_{2}$ reveals the complete structure of the TOI. The incident fields from both Tx sources are blocked by the convex part of the TOI, resulting in unilluminated regions within the domain of interest. This phenomenon can easily happen in realistic NF passive radar imaging scenarios with complex objects and geometrical arrangements. In contrast, when the contributions of the two illumination sources are combined according to \eqref{eq:sumF}, the entire TOI is accurately reconstructed, as illustrated in Fig.~\ref{fig:pyramid_comb}.

The second simulation involved a PEC $90^{\circ}$ dihedral corner reflector, as illustrated in Fig.~\ref{fig:Simu}(b). The seven positions of the illuminating dipole $\veg{r}'_{n}$ were linearly arranged in the $xz$-plane, i.e., \mbox{$\veg{r}'_{n}=[x_{n},0,0.4]\,$m} where \mbox{$x_{n}\in \{\pm0.3,\,\pm0.2,\,\pm0.1,\,0 \}$}. The frequency range was chosen from $\SI{2}{\giga\hertz}$ to $\SI{12}{\giga\hertz}$ to ensure sufficient range resolution~\cite{wang2024TAP}, with an equal step size of $\SI{250}{\mega\hertz}$. The planar measurement arrangement remained unchanged.
% \begin{figure}[t]
% 	\centering
% 	\includegraphics[scale=0.45]{Figures/Corner.pdf}
% 	\caption{Illustration of the simulation setup with a dihedral corner reflector. Simulations were performed separately for seven different positions from $\veg{r}'_{1}$ to $\veg{r}'_{7}$ of the Hertzian dipole serving as the Tx source.}
% 	\label{fig:Corner}
% \end{figure}%
\begin{figure}[t]
  \centering
  \subfloat[]{\includegraphics[scale=0.63]{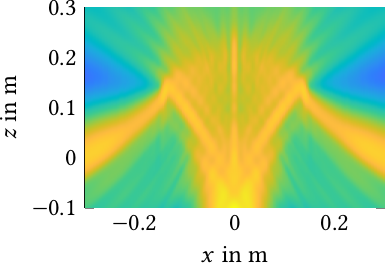}}%
  \hfill
  \subfloat[]{\includegraphics[scale=0.63]{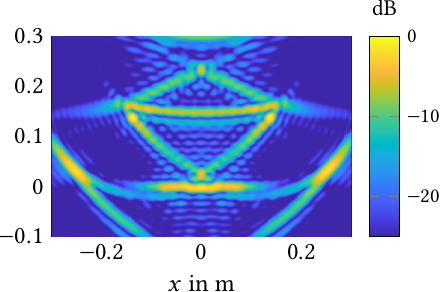}}%
  \caption{Imaging results of the dihedral corner reflector with the single illumination source located at $\veg{r}'_{4}=[0,0,0.4]\,$m. 
  (a)~Incoherent summation. 
  (b)~Coherent summation using only 11 frequencies.}
  \label{fig:corner_bad}
\end{figure}

\begin{figure}[t]
  \centering
  \subfloat[]{\includegraphics[scale=0.64]{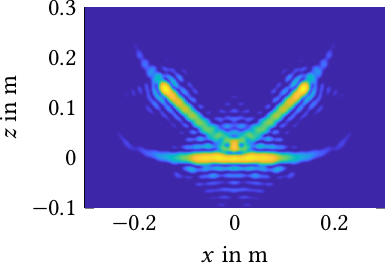}}%
  \hfill
  \subfloat[]{\includegraphics[scale=0.64]{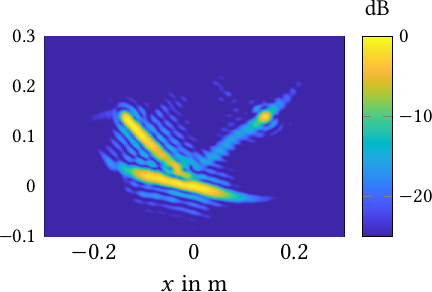}}%
  \\ 
  \vspace{-5mm}
  \subfloat[]{\includegraphics[scale=0.64]{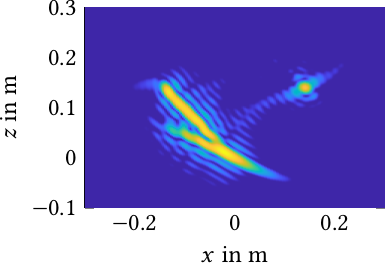}}%
   \hfill
  \subfloat[]{\includegraphics[scale=0.64]{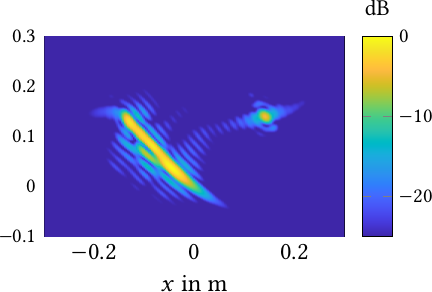}}%
  \caption{Imaging results of the dihedral corner reflector with the single illumination source located at $\veg{r}'_{4}$~(a), $\veg{r}'_{5}$~(b), $\veg{r}'_{6}$~(c), and $\veg{r}'_{7}$~(d).}
  \label{fig:corner_sep}
\end{figure}

\begin{figure}[t]
	\centering
	\includegraphics[scale=0.9]{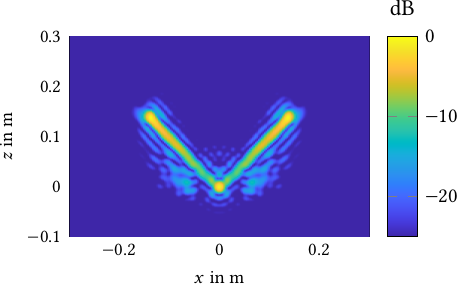}
	\caption{Coherent imaging result of the dihedral corner reflector combining the illuminating contributions of the Tx antenna moving from $\veg{r}'_{1}$ to $\veg{r}'_{7}$.}
	\label{fig:corner_comb}
\end{figure}%
% \begin{figure}[t]
%   \centering
%   \subfloat[]{\includegraphics[scale=0.8]{Figures/corner_top.pdf}}%
%   \\ 
%   \vspace{-7mm}
%   \centering
%   \subfloat[]{\includegraphics[scale=0.8]{Figures/corner_front.pdf}}%
%   \caption{Imaging result of the dihedral corner reflector combining the illuminating contributions from the Tx antenna from $\veg{r}'_{1}$ to $\veg{r}'_{7}$. 
%   (a)~Cut plane at $z=0$. 
%   (b)~MIP front view in $xy$-plane}
%   \label{fig:corner_comb}
% \end{figure}
First, the single Tx source located at $\veg{r}'_{4}=[0,0,0.4]\,$m is considered. The imaged domain is defined in the $xz$-plane. When applying incoherent summation~\cite{holl2017holography,wang2024TAP}, the resulting multi-frequency image is heavily affected by strong artifacts, as shown in Fig.~\ref{fig:corner_bad}(a). This result is improved through coherent summation, as demonstrated in Fig.~\ref{fig:corner_bad}(b), where only a subset of the single-frequency images, i.e., from $\SI{2}{\giga\hertz}$ to $\SI{12}{\giga\hertz}$ with an equal step size of $\SI{1}{\giga\hertz}$, is combined. Utilizing the full frequency range further enhances the imaging results, as shown in Fig.~\ref{fig:corner_sep}(a), where parts of the artifacts are further suppressed due to the phase correction. 

However, the phase correction term $\psi_{\mathrm{s},\,p}\left(k_f,\veg{S}_{n},\veg{r}'\right)$ accounts only for the direct propagation path and does not consider the multiple reflections from the dihedral. This limitation results in strong multipath ghosts in the image, which cannot be mitigated even with an increased frequency range. The multipath ghosts are configuration-dependent, with their positions varying based on the position of the Tx antenna, as shown in Figs.~\ref{fig:corner_sep}(b), (c), and (d). By leveraging this dependency and recognizing that the true positions of the TOIs are configuration-independent, a significantly improved result is achieved by combining the contributions from all the sources using~\eqref{eq:sumF}, as demonstrated in Fig.~\ref{fig:corner_comb}. During this process, the images of the TOIs are coherently superimposed, revealing the complete structures (including the weaker parts that were absent in the single Tx images), and the configuration-dependent multipath ghosts are greatly suppressed compared to the enhanced TOIs.

\label{sec:measurement}
\begin{figure}[t]
	\centering
  \hspace*{\fill}
	\subfloat[]{\includegraphics[scale=0.3]{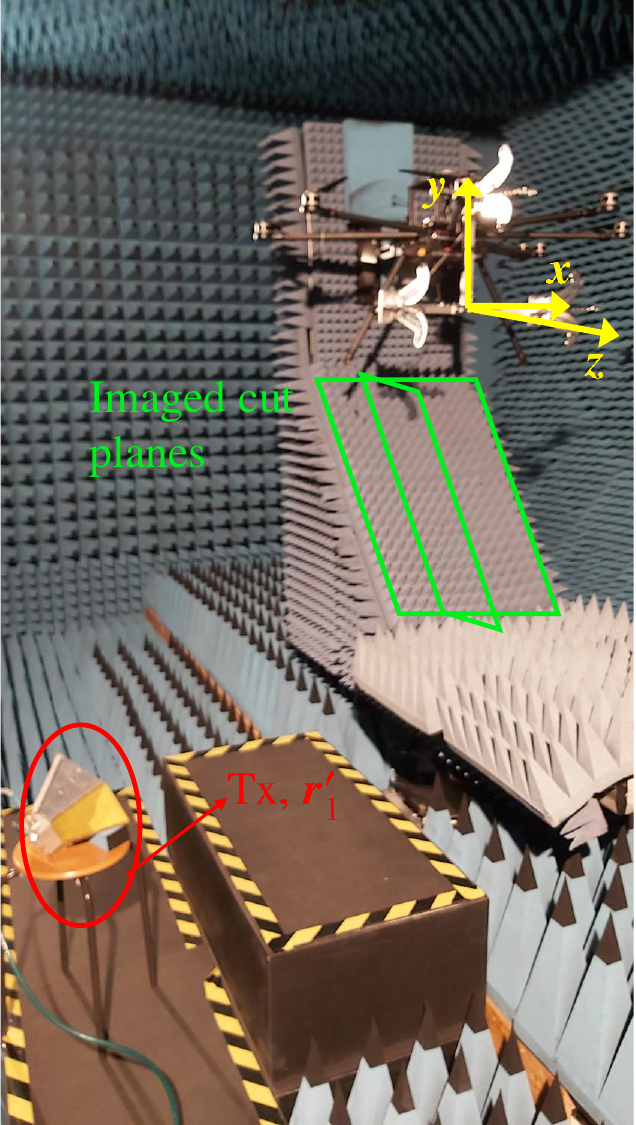}}%
	\hfill
	\subfloat[]{\includegraphics[scale=0.3]{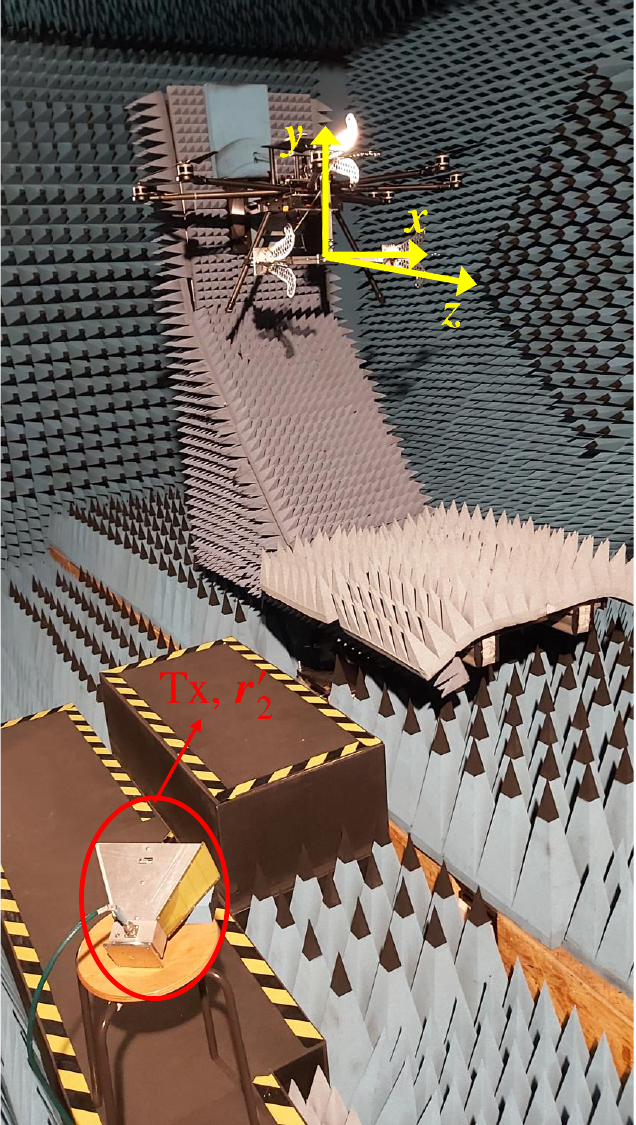}}%
  \hspace*{\fill}
	\caption{Two measurement configurations in the anechoic chamber, with the Tx antenna positioned at two different locations.
  (a)~Tx antenna at $\veg{r}'_{1}$, consistent with the setup described in \cite{wang2024TAP,wang2024microwavea}.
  (b)~Tx antenna at $\veg{r}'_{2}$, positioned slightly farther from the UAV.}
	\label{fig:Measurement}
\end{figure}

\begin{figure}[t]
	\centering
	\includegraphics[scale=0.8]{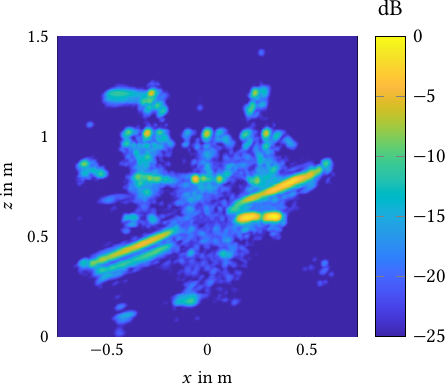}
	\caption{Imaging result of the UAV with a MIP top view in $xz$-plane, demonstrating improved quality compared to the results presented in~\cite{wang2024TAP}.}
	\label{fig:UAV}
\end{figure}%

\begin{figure}[t]
	\centering
	\subfloat[]{\includegraphics[scale=0.62]{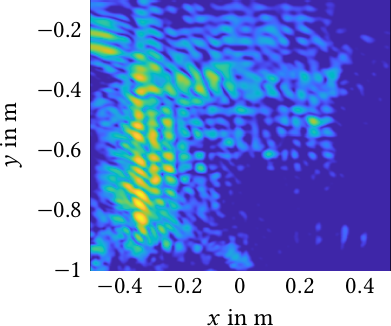}}%
	\hfill
	\subfloat[]{\includegraphics[scale=0.62]{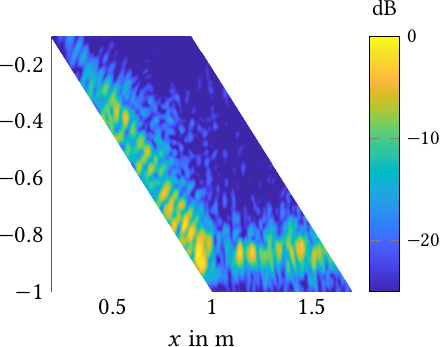}}%
	\caption{Imaging results of the absorbers, demonstrating improved quality compared to the results presented in~\cite{wang2024microwavea}.
  (a)~In the plane parallel to the slope (see Fig.~\ref{fig:Measurement}(a)) with $z = -0.9y+0.2\,$m.
  (b)~In the plane perpendicular to the slope with $x=\SI{-0.33}{\meter}$.}
	\label{fig:absorber}
\end{figure}
% \balance

Furthermore, a measurement campaign was conducted in an anechoic antenna measurement chamber. The measurement setup depicted in Fig.~\ref{fig:Measurement}(a) and the corresponding single Tx imaging results for an uninhabited aerial vehicle (UAV) and the microwave absorbers connected to the positioner have been detailed in~\cite{wang2024TAP} and~\cite{wang2024microwavea}. Multiple sources measurements were performed by moving the Tx horn antenna further away from the UAV, while all other parameters, as described in~\cite{wang2024TAP,wang2024microwavea}, remained unchanged, as shown in Fig.~\ref{fig:Measurement}(b). 

The Tx source positions determined from the phase correction processes are $\veg{r}'_{1}=[-1.46,-1.06, 1.31]\,$m~\cite{wang2024TAP} and $\veg{r}'_{2}=[-1.48,-1.06,2.35]\,$m, respectively. The 2-D MIP top view of the imaged UAV and the absorbers in two cut planes, similar to those presented in~\cite{wang2024TAP,wang2024microwavea} are shown in Fig.~\ref{fig:UAV} and Fig.~\ref{fig:absorber}, respectively. By combining the imaging results from multiple sources, Fig.~\ref{fig:UAV} and Fig.~\ref{fig:absorber} demonstrate improved artifact suppression. Additionally, sharper features of the TOIs, such as the horizontal part of the absorbers in Fig.~\ref{fig:absorber}(b) are revealed more clearly.

\section{Conclusion}
\label{sec:conclusion}
An enhanced inverse source based passive radar imaging method utilizing multiple illumination sources was presented. For each illumination source, single-frequency images are generated through inverse source reconstruction and are coherently superimposed using the phase and magnitude correction methods. The omission of multiple reflections in the analysis leads to the presence of strong multipath ghosts in the obtained multi-frequency images. Leveraging the configuration-dependent nature of these multipath ghosts and combining the contributions of multiple sources effectively suppresses them. Moreover, the use of multiple illumination sources reduces the likelihood of unilluminated regions in complex NF passive imaging scenarios. Overall, this approach enables application to a broader range of scenarios and achieves superior imaging performance.

% use section* for acknowledgement
\section*{ACKNOWLEDGEMENT}
Funded by the European Union. Views and opinions expressed are however those of the author(s) only and do not necessarily reflect those of the European Union or European Innovation Council and SMEs Executive Agency (EISMEA). Neither the European Union nor the granting authority can be held responsible for them. Grant Agreement No: 101099491.

\bibliographystyle{IEEEtran}
\bibliography{Literature.bib}

\end{document}